\documentclass[journal]{IEEEtran}
\IEEEoverridecommandlockouts
\usepackage{cite}
\usepackage{amsmath,amssymb,amsfonts}
\usepackage{algorithm}
\usepackage{algpseudocode}
\usepackage{graphicx}
\usepackage[hidelinks]{hyperref}
\usepackage{multirow}
\usepackage{booktabs}
\usepackage{textcomp}
\usepackage{xcolor}
\usepackage{tabularx}
\usepackage{ragged2e}
\usepackage[english]{babel}
\usepackage{stfloats}
\def\BibTeX{{\rm B\kern-.05em{\sc i\kern-.025em b}\kern-.08em
    T\kern-.1667em\lower.7ex\hbox{E}\kern-.125emX}}
\newcolumntype{A}{>{\raggedright\arraybackslash}p{2.4cm}}
\newcolumntype{B}{>{\RaggedRight\arraybackslash}X}

\begin{document}

\title{Benchmarking Cyberattack Detection in Electric Vehicle Charging Infrastructure with User Updates}

\author{Hannan~Chen,~\IEEEmembership{Graduate Student Member,~IEEE},
Roshni~Anna~Jacob,~\IEEEmembership{Member,~IEEE},
and~Jie~Zhang,~\IEEEmembership{Senior Member,~IEEE}%
\thanks{Hannan Chen and Roshni Anna Jacob are with the Department of Mechanical Engineering, The
University of Texas at Dallas, Richardson, TX 75080, USA.}%
\thanks{Jie Zhang is with the Department of Mechanical Engineering and (Affilated) Department of Electrical and Computer Engineering, The University of Texas at Dallas, Richardson, TX 75080, USA (e-mail: jiezhang@utdallas.edu).}}

\maketitle

\begin{abstract}
Cyberattack detection in electric vehicle charging infrastructure is complicated by legitimate post-activation revisions to requested energy and departure time. Charging manipulation attacks can exploit the same interface and variables; therefore, detecting a request change alone does not establish malicious intent. This paper develops a leakage-controlled session-level benchmark that preserves the ordered inputs of real Adaptive Charging Network (ACN) sessions and models legitimate revisions as normal behavior. A fixed pool keeps each generated attack in its source session's split and contains six physically motivated attacks and their coordinated variants. We compare 22 profile-only, transition-aware, and context-stratified model families under common source-grouped folds, attack data, and operating constraints. The proposed Dual-Branch Masked-Autoencoder (Masked-AE) Transition Boost model evaluates whether the current request is normal and whether its producing transition resembles an observed benign update. Its state branch combines masked reconstruction with a radial-basis-function one-class support boundary, while its transition branch combines masked reconstruction with shrinkage covariance distance. Source-grouped five-fold cross-validation selects complete configurations under explicit overall-normal and benign-update acceptance constraints; disjoint normal data then calibrate the final threshold before one test evaluation. The developed dual-branch model provides the strongest robust validation performance while detecting malicious request manipulations without learning to reject legitimate user choices.
\end{abstract}


\section{Introduction}
The growing adoption of electric vehicles (EVs) introduces critical cybersecurity vulnerabilities to the power grid, highlighting the vital role of secure smart charging infrastructure. Smart charging coordinates users, charging points, charging-station management systems (CSMSs), and grid operators to optimize charging times and rates according to mobility requests, grid conditions, and electricity prices. Charging stations exchange requested energy, connection time, departure time, and operating status with back-end systems, commonly through the Open Charge Point Protocol (OCPP)~\cite{gridx_smart_charging}.

This connectivity also enlarges the cyber-physical attack surface. User applications, charging stations, and back ends can be exposed to message manipulation and man-in-the-middle attacks~\cite{9800931,alcaraz2017ocpp_tsg}. Secure channels and authentication are necessary, but a compromised authorized component can still submit syntactically valid yet malicious charging requests. Such requests can increase peak demand, shift aggregate load, interfere with demand response, and affect grid operation. Runtime behavioral detection is therefore required in addition to communication security.

\subsection{Literature Review}
Smart EV charging depends on continuous communication between EV supply equipment and CSMSs. Prior OCPP studies identify message manipulation, fraudulent transactions, reservation or transaction interference, and man-in-the-middle threats~\cite{alcaraz2017ocpp_tsg}. Secure channel establishment and stronger authentication and authorization reduce these risks~\cite{rubio2018ocpp_ntms}; nevertheless, practical demonstrations of denial-of-service, man-in-the-middle, and code-injection exploits show why operational monitoring remains necessary~\cite{elmo2023ocpp_rws}.

Within this ecosystem, charging manipulation attacks (CMAs) are especially consequential because they alter requested start time, expected departure, and/or requested energy to reshape aggregate demand. Jahangir \emph{et al.} ~\cite{10531054} formalized CMAs against smart EV charging stations and analyzed how coordinated request manipulation can shift an aggregator's load across the day. They also evaluated deep-learning detection mechanisms on real charging-session data, establishing both a physically motivated threat model and an important data-driven baseline. 

The present study retains that central attack logic and its peak-shifting objective, but makes the normal-data model more realistic: every qualifying ordered user input is preserved, attacks may originate from the request state currently in effect, and a detector must distinguish malicious changes from recorded benign energy and departure revisions.

In practice, labeled field attacks are scarce, while normal charging is heterogeneous. A variety of methods have been explored in the literature to address such challenges~\cite{11225416,10935626,10402098,liu2026gdgugradientdifferencebasedgraph,11319260}. Autoencoders (AEs) learn nonlinear normal representations through reconstruction~\cite{hinton2006ae_science,sakurada2014ae_anomaly}; denoising objectives infer corrupted features from the remaining coordinates~\cite{vincent2008denoising}, motivating the masked AEs used here. Gaussian Error Linear Unit (GELU) activations smoothly process standardized continuous requests~\cite{hendrycks2016gelu}. One-class support vector machines (OCSVMs) estimate nonlinear normal support~\cite{scholkopf2001estimating}, while local outlier factor (LOF), Gaussian mixture model (GMM), $k$-nearest-neighbor (KNN), and covariance heads test local, density, and distance assumptions~\cite{10.1145/342009.335388}. Deep Support Vector Data Description (SVDD), deviation networks (DevNet), and real-valued non-volume-preserving (RealNVP) normalizing flows (NFs) provide neural center, deviation, and invertible-density alternatives~\cite{ruff2018deep,pang2019devnet,dinh2016realnvp}.

The preceding methods play two roles in this study. Masked AEs, the RBF OCSVM, and shrinkage covariance estimation form the proposed detector, whereas LOF, GMM, KNN, Deep SVDD, DevNet, and RealNVP NF models serve as benchmark comparators. In particular, the proposed detector in this paper contains no NF block. Existing single-profile detectors generally assess the current request without separately modeling the transition that produced it, although benign updates and CMAs modify the same demand and timing variables.

To address this gap, Dual-Branch Masked-AE Transition Boost is developed to evaluate two complementary forms of evidence. For every activation or update, its state branch uses a masked AE and an RBF OCSVM to determine whether the current request lies within normal support. For post-activation updates, its transition branch uses a second masked AE and Ledoit--Wolf covariance distance to determine whether the change from the preceding request resembles an observed benign update~\cite{ledoit2004wellconditioned}. Standardized branch scores are combined through an update-gated, low-weight transition correction, while activation events remain state-only. Thus, an abnormal transition can promote an individually plausible attack, whereas a normal transition can reduce suspicion for a rare legitimate revision.

\subsection{Contributions}
This work makes four contributions:
\begin{itemize}
\item It formulates multi-input CMA detection as joint assessment of the current request and the transition that produced it, making correct acceptance of recorded benign updates an explicit objective rather than treating every post-activation change as anomalous.
\item It constructs a source-bound benchmark from all qualifying ordered Adaptive Charging Network (ACN) inputs, retains recorded energy and departure updates as normal behavior, reproduces their piecewise scheduled-power effects, and applies six physical CMA transformations and six coordinated variants once to eligible request states.
\item It develops Dual-Branch Masked-AE Transition Boost, whose state branch learns pooled normal-request support and whose transition branch learns the geometry of observed normal updates; gated continuous fusion preserves sensitivity to attacks while allowing legitimate revisions.
\item It evaluates 89 profile-only, transition-aware, and context-stratified configurations using source-grouped five-fold cross-validation (CV), equal-weight attack-family scoring, and explicit overall-normal and benign-update acceptance constraints before final testing.
\end{itemize}

The remainder of this paper is organized as follows. Section II presents the multi-input data and CMA generation. Section III develops the unsupervised models and evaluation protocol. Section IV reports the results, followed by discussion, limitations, and conclusions.

\section{Cyberattack Scenario Generation}
\subsection{Multi-Input ACN Data}
The study uses 1,213 valid Caltech ACN sessions~\cite{lee_acndata_2019}. Each record includes connection and disconnect times, station metadata, and an ordered `userInputs' list. Let $u_{n,\ell}$ be input $\ell$ of session $n$. Input zero defines activation. For $\ell>0$, an observed update is retained when
\begin{equation}
c_{n,\ell}=\mathbf 1\!\left(|d_{n,\ell}-d_{n,\ell-1}|>\epsilon_{\mathrm{chg}}
\ \lor\ et_{n,\ell}\ne et_{n,\ell-1}\right).
\label{eq:update_indicator}
\end{equation}
Here $\epsilon_{\mathrm{chg}}$ is a fixed numerical tolerance that removes no-op energy records.
Its event time and request state are:
\begin{equation}
\begin{aligned}
t^{\mathrm{in}}_{n,\ell}&=\mathrm{modifiedAt}(u_{n,\ell}),\\
st_n&=\mathrm{modifiedAt}(u_{n,0}),\\
et_{n,\ell}&=\mathrm{requestedDeparture}(u_{n,\ell}),\\
d_{n,\ell}&=\mathrm{kWhRequested}(u_{n,\ell}).
\end{aligned}
\end{equation}
Among the 1,213 valid sessions, 73 contain more than one `userInputs' record. Duplicate/no-op records are not events; only 45 sessions contain a qualifying consecutive change in the two studied decision variables. The resulting 1,264 normal states comprise 1,213 activations and 51 updates: 13 demand-only, 14 departure-only, and 24 joint updates.

For example, one ordinary initial input requests 38.87 kWh from 14:29 to 22:29, corresponding to an implied rate of 4.86 kW. Table~\ref{tab:update_examples} gives three observed consecutive updates. It shows that benign revisions can move in either direction and can be substantial; these values are read directly from ACN `userInputs', not synthetically perturbed.

\begin{table}[thb!]
\centering
\caption{Examples of observed benign user updates. Times are local.}
\label{tab:update_examples}
\scriptsize\renewcommand{\arraystretch}{1.08}\setlength{\tabcolsep}{3pt}
\begin{tabular}{llll}
\toprule
\textbf{Source ID} & \textbf{Kind} & \textbf{Energy (kWh)} & \textbf{Departure}\\
\midrule
75 & Demand & $68.0\rightarrow34.0$ ($-50\%$) & 18:49 (unchanged)\\
1108 & Departure & 35.0 (unchanged) & $10{:}58\rightarrow11{:}58$\\
546 & Joint & $28.6\rightarrow42.9$ ($+50\%$) & $11{:}47\rightarrow12{:}09$\\
\bottomrule
\end{tabular}
\end{table}

\noindent The five request features are:
\begin{equation}
\begin{aligned}
T_{n,\ell}&=(et_{n,\ell}-st_n)/3600,\qquad
Ch_{n,\ell}=d_{n,\ell}/T_{n,\ell},\\
x_{n,\ell}&=[d_{n,\ell},Ch_{n,\ell},st_n-t_{\mathrm{peak}},
et_{n,\ell}-t_{\mathrm{peak}},T_{n,\ell}]^\top,
\end{aligned}
\label{eq:feature_vector}
\end{equation}
where $T_{n,\ell}$ and $Ch_{n,\ell}$ are request duration and implied rate, and $t_{\mathrm{peak}}=18{:}30$ local time. Thus, $t^{\mathrm{in}}_{n,\ell}$ triggers a state change but never restarts charging: every observed input retains $st_n$. In each fold, feature means and standard deviations estimated only from training-normal states are applied unchanged to all downstream data. All states and attacks from a source remain in one split and one fold.

\subsection{Charging Manipulation Attacks}
The six CMAs increase or concentrate charging in the 16:00--21:00 peak interval. For an ordinary Type-$i$ attack on session $n$, the attack is launched at $ta=st_n+dt$, where $dt=5$ min, and changes one eligible request state. The suffix ``-ta'' denotes the coordinated version, which applies the same transformation to multiple states active at a common $ta$. The attack charging rate (ACR) is:
\begin{equation}
ACR_{\mathrm{att}}=C_{\mathrm{att}}Ch_{\mathrm{av}}=0.15(30)=4.5\ \mathrm{kW}.
\label{eq:acr}
\end{equation}
Here $C_{\mathrm{att}}$ is the attack coefficient and $Ch_{\mathrm{av}}$ is the reference charging rate. Their product balances physical impact and stealth: it changes aggregate load while keeping many individual requests plausible and below the simulator's per-session power limit $P_l=24$ kW. For an attacked source, define the input active immediately before the attack as:
\[
\ell_n(ta)=\max\{\ell:t^{\mathrm{in}}_{n,\ell}\le ta\},
\]
and abbreviate its request state by $d_n(ta^-)=d_{n,\ell_n(ta)}$,
$et_n(ta^-)=et_{n,\ell_n(ta)}$, and
$Ch_n(ta^-)=Ch_{n,\ell_n(ta)}$. With time differences expressed in hours,
\begin{equation}
\begin{aligned}
H_n(ta)&=et_n(ta^-)-ta,\\
E_n^{\mathrm{rem}}(ta)&=Ch_n(ta^-)H_n(ta),\\
\Delta d_i^{\mathrm{att}}(n,ta)&=\xi_iACR_{\mathrm{att}}H_n(ta).
\end{aligned}
\label{eq:attack_state}
\end{equation}
These quantities are the remaining time, remaining requested energy, and Type-$i$ demand increment.
Table~\ref{tab:cma_types} gives the exact transformations; omitted settings are unchanged.

\begin{table*}[thb!]
\centering
\caption{Exact transformations for Type-1--6; each is also used in Type-$i$-ta.}
\label{tab:cma_types}
\scriptsize\renewcommand{\arraystretch}{1.08}\setlength{\tabcolsep}{3pt}
\begin{tabular}{p{.8cm}p{2.7cm}p{13cm}}
\toprule
\textbf{Type} & \textbf{Manipulation} & \textbf{Transformed settings}\\
\midrule
1 & Demand & $\overline d^{n,1}_{ta}=d_n(ta^-)+\Delta d_1^{\mathrm{att}}(n,ta),\quad \overline{st}^{n,1}_{ta}=st_n,\quad \overline{et}^{n,1}_{ta}=et_n(ta^-)$.\\
2 & Earlier departure & $\overline{Ch}^{n,2}_{ta}=Ch_n(ta^-)+\xi_2ACR_{\mathrm{att}},\quad \overline T^{n,2}_{ta}=E_n^{\mathrm{rem}}(ta)/\overline{Ch}^{n,2}_{ta},\quad \overline{et}^{n,2}_{ta}=ta+\overline T^{n,2}_{ta},\quad \overline d^{n,2}_{ta}=d_n(ta^-)$.\\
3 & Later start & $\overline T^{n,3}_{ta}=E_n^{\mathrm{rem}}(ta)/(Ch_n(ta^-)+\xi_3ACR_{\mathrm{att}}),\quad \overline{st}^{n,3}_{ta}=et_n(ta^-)-\overline T^{n,3}_{ta},\quad \overline d^{n,3}_{ta}=d_n(ta^-)$.\\
4 & Demand + earlier departure & $\overline d^{n,4}_{ta}=d_n(ta^-)+\Delta d_4^{\mathrm{att}}(n,ta),\quad \widetilde T^{n,4}_{ta}=[E_n^{\mathrm{rem}}(ta)+\Delta d_4^{\mathrm{att}}(n,ta)]/[Ch_n(ta^-)+ACR_{\mathrm{att}}],\quad \overline{et}^{n,4}_{ta}=ta+\widetilde T^{n,4}_{ta}$.\\
5 & Demand + later start & $\overline d^{n,5}_{ta}=d_n(ta^-)+\Delta d_5^{\mathrm{att}}(n,ta),\quad \widetilde T^{n,5}_{ta}=[E_n^{\mathrm{rem}}(ta)+\Delta d_5^{\mathrm{att}}(n,ta)]/[Ch_n(ta^-)+ACR_{\mathrm{att}}],\quad \overline{st}^{n,5}_{ta}=et_n(ta^-)-\widetilde T^{n,5}_{ta}$.\\
6 & Demand + both times & $\overline d^{n,6}_{ta}=d_n(ta^-)+\Delta d_6^{\mathrm{att}}(n,ta),\quad \widetilde T^{n,6}_{ta}=[E_n^{\mathrm{rem}}(ta)+\Delta d_6^{\mathrm{att}}(n,ta)]/[Ch_n(ta^-)+ACR_{\mathrm{att}}],\quad \Delta T^{n,6}_{ta}=H_n(ta)-\widetilde T^{n,6}_{ta},\quad \overline{et}^{n,6}_{ta}=et_n(ta^-)-\kappa_n\Delta T^{n,6}_{ta},\quad \overline{st}^{n,6}_{ta}=ta+(1-\kappa_n)\Delta T^{n,6}_{ta}$.\\
\bottomrule
\end{tabular}
\end{table*}

\begin{figure*}[thb!]
\centering
\includegraphics[width=.9\textwidth]{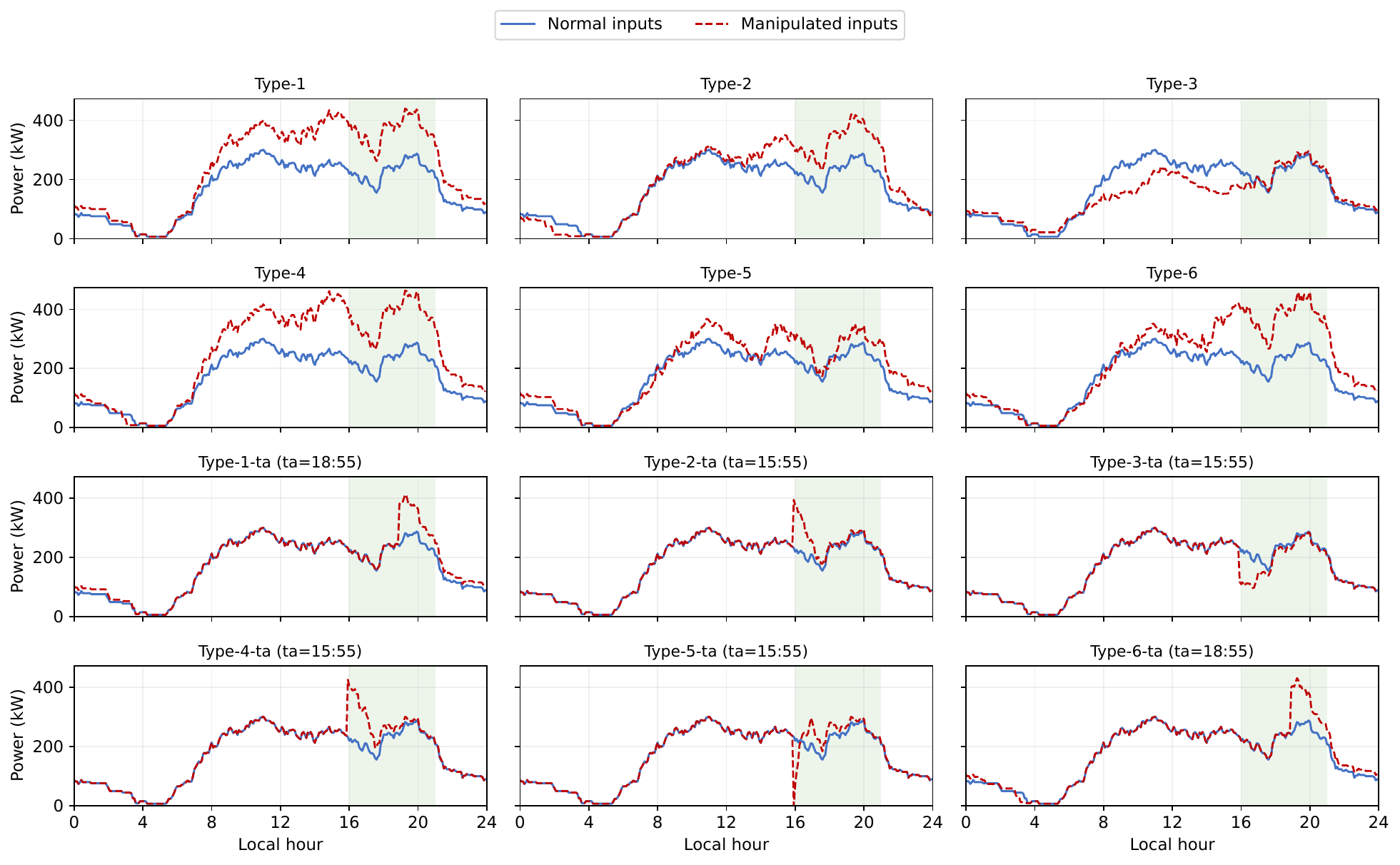}
\caption{Aggregate charging profiles for the twelve fixed CMA settings. Power follows Table~\ref{tab:cma_types} and~\eqref{eq:initial_power}--\eqref{eq:piecewise_power}; shading marks the 16:00--21:00 peak-price period.}
\label{fig:cma_profiles}
\end{figure*}

Types 1--3 isolate demand, departure, and start-time changes; Types 4--6 combine demand growth with time compression. The factor $\xi_i\in(0,1)$ randomizes strength, with standard deviation 0.1 and means 0.9 for Types 1--3 and 0.8 for Types 4--6; $\kappa_n$ allocates the Type-6 reduction between start delay and departure advance. Algorithm~\ref{alg:attack_pool} summarizes the one-time, source-bound generation with seed 66. CV attacks originate only from validation sources and final attacks only from final-test sources. Figure~\ref{fig:cma_profiles} shows the fixed aggregate effects.

\begin{algorithm}[thb!]
\caption{Source-bound multi-input CMA pool}
\label{alg:attack_pool}
\begin{algorithmic}[1]
\Require Ordered normal inputs, source roles, CMA controls, seed 66.
\For{each validation or final-test role}
 \For{each eligible input state and Type-1--6/single-coordinated mode}
  \State Apply Table~\ref{tab:cma_types}; recompute~\eqref{eq:feature_vector}.
  \State Save attack, source, input, type, and role identifiers.
 \EndFor
\EndFor
\State Reuse the saved pool for every detector configuration.
\end{algorithmic}
\end{algorithm}

\subsection{Observed Benign Updates}
Observed updates are normal because they are exact successive ACN inputs, not generated perturbations. At $t^{\mathrm{in}}_{n,\ell}$, the recorded $d_{n,\ell}$ and $et_{n,\ell}$ replace the current request while $st_n$ remains fixed. Each \texttt{kWhRequested} is a revised total target, not an added increment. Equation~\eqref{eq:feature_vector} recomputes the request features. Physical power follows the implemented piecewise charging semantics rather than the implied-rate feature. At activation $t_0=st_n$,
\begin{align}
P_n(t_0^+)&=\frac{d_{n,0}}{(et_{n,0}-st_n)/3600},\qquad
E_n^{\mathrm{del}}(t_0)=0,\nonumber\\
b_n(t_0^+)&=t_0,\qquad e_n(t_0^+)=\min(et_{n,0},disc_n).
\label{eq:initial_power}
\end{align}
Here $P_n$ is scheduled power, $E_n^{\mathrm{del}}$ is cumulative delivered energy, $b_n$ and $e_n$ bound a power segment, and $disc_n$ is the recorded disconnect time.
For a later event at $t$, let $b_n(t^-)$ and $e_n(t^-)$ delimit the preceding constant-power segment. The delivered energy and new segment are:
\begin{align}
E_n^{\mathrm{del}}(t)&=E_n^{\mathrm{del}}(b_n(t^-))+P_n(t^-)\nonumber\\
&\quad\times\frac{[\min(t,e_n(t^-))-b_n(t^-)]_+}{3600},\nonumber\\
P_n(t^+)&=\frac{d_n(t^+)-E_n^{\mathrm{del}}(t)}{[et_n(t^+)-t]/3600},\nonumber\\
b_n(t^+)&=t,\nonumber\\
e_n(t^+)&=\min(et_n(t^+),disc_n).
\label{eq:piecewise_power}
\end{align}
Thus, energy delivered before an attack or user update is retained; \textit{disconnect} truncates the active interval but does not replace the requested departure in the power denominator. Benign and malicious events differ by their parameter updates, not by power accounting. ACN does not guarantee a complete interface-level edit history; the 51 events are therefore an observed subset, not a population-rate estimate.

\section{Detection Methodology}
\subsection{Multi-Input Unsupervised Protocols}
\begin{figure*}[thb!]
\centering
\includegraphics[width=.7\textwidth]{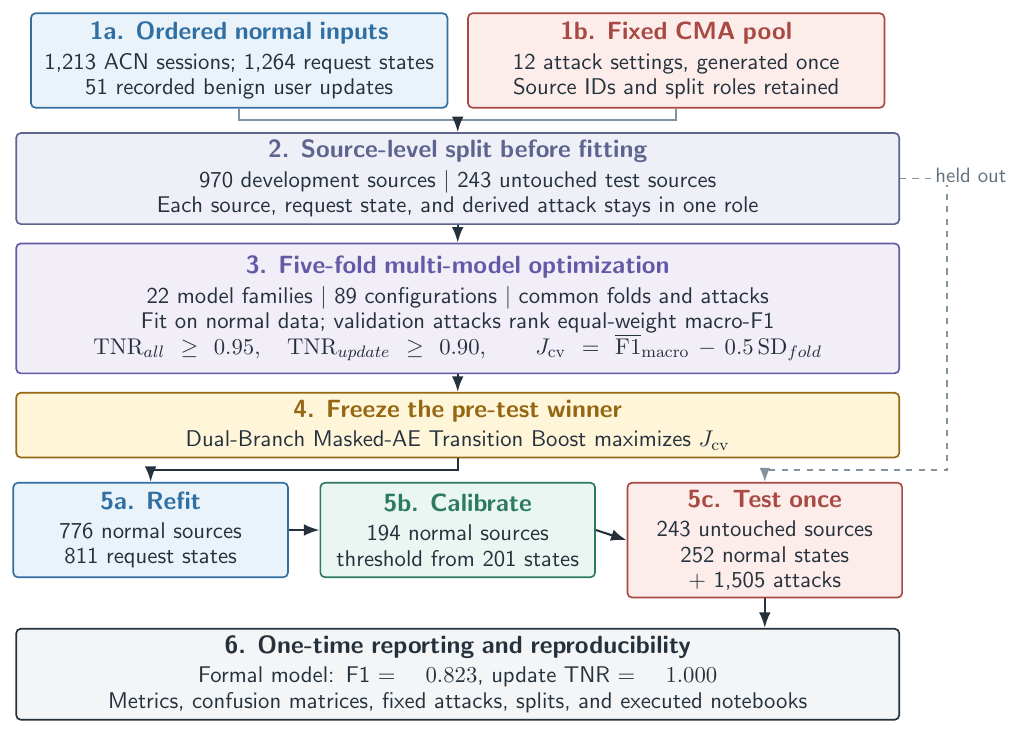}
\caption{Workflow from ordered ACN inputs and a source-bound CMA pool to source-grouped selection, normal-only calibration, and final testing.}
\label{fig:protocol}
\end{figure*}

Figure~\ref{fig:protocol} summarizes the leakage-controlled experiment from ordered ACN inputs to final testing. The three protocols answer progressively different questions. Profile-only models ask whether the current request itself is normal and use $X^{\mathrm{prof}}_{n,\ell}=x_{n,\ell}$, pooling activations with updates. Transition-aware models additionally ask how the request changed and use:
\begin{equation}
X^{\mathrm{tr}}_{n,\ell}=[x_{n,\ell}^\top,\Delta x_{n,\ell}^\top,c_{n,\ell}]^\top,
\quad \Delta x_{n,\ell}=x_{n,\ell}-x_{n,\ell-1},
\label{eq:transition_features}
\end{equation}
with zero change for activation. Context-stratified models instead learn separate normal models for activations and updates, then put their scores on a common scale before applying one threshold. These protocols test whether explicit update context helps or whether scarce update samples make a pooled current-state model more stable.

Normal-only means that no CMA sample estimates model parameters; simulated validation attacks only select among predefined fitted configurations. DevNet uses artificial tail samples as pseudo-anomalies, but not the fixed benchmark attacks. A model predicts
\begin{equation}
\widehat y(x)=\mathbf 1[s(x)\ge s_{\mathrm{thr}}].
\label{eq:threshold_rule}
\end{equation}
We denote the training, validation, normal-only calibration, and final-test sets by $\mathcal D_{\mathrm{tr}}$, $\mathcal D_{\mathrm{val}}$, $\mathcal D_{\mathrm{cal}}$, and $\mathcal D_{\mathrm{te}}$, respectively.

\subsection{Model Families}
For AE models, encoder $f_\theta$ and decoder $g_\theta$ minimize normal reconstruction error:
\begin{equation}
\begin{aligned}
z&=f_\theta(\widetilde x),& \widehat x&=g_\theta(z),\\
\varepsilon_{\mathrm{AE}}&=\widetilde x-\widehat x,&
\mathcal L_{\mathrm{rec}}&=\|\varepsilon_{\mathrm{AE}}\|_2^2,\\
h_{\mathrm{AE}}&=[\varepsilon_{\mathrm{AE}}^\top,z^\top]^\top.&&
\end{aligned}
\label{eq:ae}
\end{equation}
The downstream head evaluates latent rarity and signed residuals. To learn conditional feature relationships, a masked AE draws $M_p\sim\mathrm{Bernoulli}(\rho)$, sets masked coordinates to zero, and minimizes
\begin{equation}
\mathcal L_{\mathrm{MAE}}=
\frac{\sum_{p=1}^{D_x}M_p(\widehat x^{(M)}_p-\widetilde x_p)^2}
{\max(\sum_{p=1}^{D_x}M_p,1)}
+\frac{0.1}{D_x}\|\widehat x^{(M)}-\widetilde x\|_2^2.
\label{eq:masked_ae}
\end{equation}
where $D_x$ is the input dimension and $\rho$ is the masking probability.
Here $M_p=1$ hides feature $p$ but retains it as a target. The residual records feature-wise error, while the latent code locates the request in the learned normal representation.
Its state representation is $h_{\mathrm{AE}}^{\mathrm{state}}=[\varepsilon_{\mathrm{AE}}^\top,z^\top]^\top$. For an update, the dimensionless compact transition is
\begin{equation}
\delta_{n,\ell}=
\begin{bmatrix}
\Delta d/\max(|d^-|,\epsilon_{\mathrm{chg}})\\
\Delta Ch/\max(|Ch^-|,\epsilon_{\mathrm{chg}})\\
\Delta st/\max(3600|T^-|,\epsilon_{\mathrm{chg}})\\
\Delta et/\max(3600|T^-|,\epsilon_{\mathrm{chg}})\\
\Delta T/\max(|T^-|,\epsilon_{\mathrm{chg}})
\end{bmatrix},
\label{eq:compact_transition}
\end{equation}
where each difference is current minus preceding input and the factor 3600 converts $T^-$ from hours to seconds for the two timestamp coordinates. For activation, $\delta_{n,0}=0$. For an observed update, the vector compares two successive recorded inputs; for a generated attack, it compares the manipulated request with the legitimate state in effect immediately before the attack. The dual-branch candidate fits a masked-AE+OCSVM state head and a masked-AE covariance transition head on normal data. After standardizing each raw score against its training reference, it uses
\begin{equation}
s_{\mathrm{dual}}(n,\ell)=s_{\mathrm{state}}(n,\ell)+
\omega_{\mathrm{tr}}I^{\mathrm{chg}}_{n,\ell}s_{\mathrm{tr}}(n,\ell).
\label{eq:dual_score}
\end{equation}
Here $I^{\mathrm{chg}}_{n,\ell}=0$ for activation and one for an observed update or generated attack. Thus, activations use the state boundary alone, whereas an actual request change contributes graded transition evidence. The gate does not exempt user updates from detection: every event retains its state score, and only events with a genuine change receive the additional transition correction.

\begin{table*}[thb!]
\centering
\caption{Unified unsupervised model families and anomaly scores.}
\label{tab:model_families}
\footnotesize\renewcommand{\arraystretch}{1.07}\setlength{\tabcolsep}{3.2pt}
\begin{tabular}{p{3.2cm}p{6.0cm}p{7.4cm}}
\toprule
\textbf{Family} & \textbf{Normal model} & \textbf{Anomaly score}\\
\midrule
Raw support/density/distance & RBF OCSVM, KNN, Ledoit--Wolf covariance, GMM, or LOF on $X^{\mathrm{prof}}$ or $X^{\mathrm{tr}}$ & Boundary distance, neighbor distance, squared Mahalanobis distance, negative log likelihood, or local-density deficit.\\
AE heads & AE residual-latent feature $h_{\mathrm{AE}}=[\varepsilon_{\mathrm{AE}},z]$ with Mahalanobis or GMM heads & Covariance distance or mixture negative log likelihood.\\
AE+NF & RealNVP map $F_\phi$ on $h_{\mathrm{AE}}$ & $-\log p_\phi(h_{\mathrm{AE}})$, where $\log p_\phi(h_{\mathrm{AE}})=\log p_0(F_\phi(h_{\mathrm{AE}}))+\log|\det \partial F_\phi/\partial h_{\mathrm{AE}}|$~\cite{dinh2016realnvp}.\\
Masked-AE models & Masked reconstruction followed by an OCSVM support head & Boundary distance in the learned residual-latent representation.\\
Transition boosts & Raw or masked-AE state head plus a normal-only transition head & $s_{\mathrm{state}}+\omega_{\mathrm{tr}}I^{\mathrm{chg}}s_{\mathrm{tr}}$ as in Eq. \eqref{eq:dual_score}.\\
Context-conditioned models & Separate activation/update heads or station-shrunk masked-AE features & Head-standardized anomaly evidence.\\
Deep SVDD / DevNet & Centered neural embedding or neural deviation from pseudo-anomaly tails~\cite{ruff2018deep,pang2019devnet} & Center distance or standardized positive deviation.\\
\bottomrule
\end{tabular}
\end{table*}

Support, density, and distance methods test normal-region membership, probability, and distance, respectively. The 89 predefined configurations share source folds, attacks, constraints, and selection score. Table~\ref{tab:model_families} presents the principal families; the full registry is released with the reproducibility artifacts.

\subsection{Proposed Dual-Branch Masked-AE Transition Boost}
\label{sec:dual_branch}
\begin{figure*}[thb!]
\centering
\includegraphics[width=.68\textwidth] 
{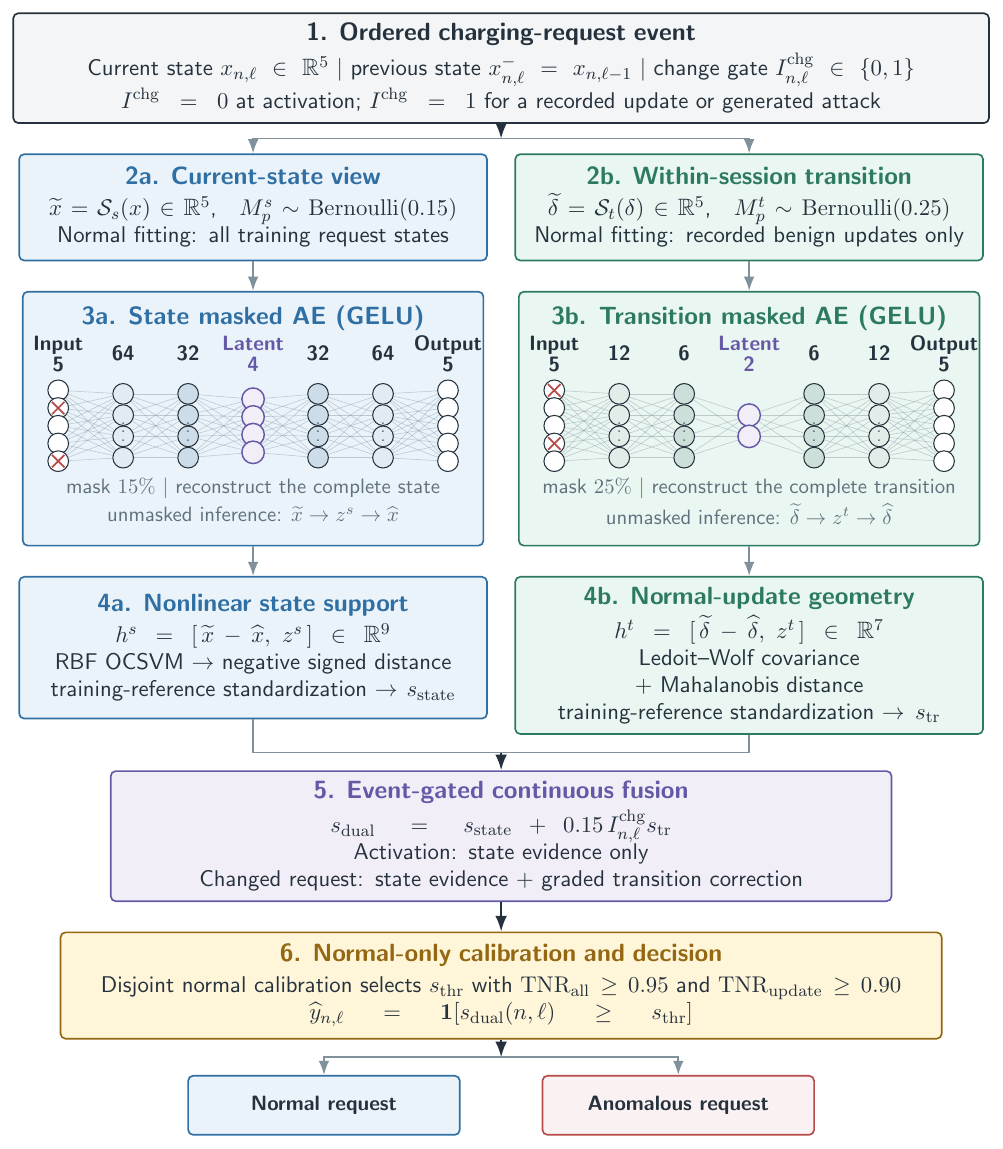}
\caption{Proposed Dual-Branch Masked-AE Transition Boost architecture. Circles show explicit input, latent, and output nodes; representative hidden nodes are shown with the exact implemented widths above each layer. The state branch learns current-request support, the transition branch learns normal user-update geometry, and the gate activates transition evidence only after a request change.}
\label{fig:dual_branch_architecture}
\end{figure*}

The proposed model separates two questions that a single pooled representation cannot answer reliably:
\emph{Is the current request state normal?} and \emph{Is the change that produced it normal?}
Figure~\ref{fig:dual_branch_architecture} follows the data path. The state branch fits all training-normal states; the transition branch fits only consecutive observed benign updates. Generated attacks enter no standardizer, AE, one-class head, or score-reference moment; they only compare configurations on validation folds and report final performance.

\subsubsection{State masked autoencoder}
Training-normal current states are standardized by a scaler $\mathcal S_s$ fitted only on $\mathcal D_{\mathrm{tr}}$:
\begin{equation}
\widetilde x_{n,\ell}=\mathcal S_s(x_{n,\ell})\in\mathbb R^5.
\label{eq:state_scale}
\end{equation}
For every minibatch, the binary mask
$M^s_{n,\ell,p}\sim\operatorname{Bernoulli}(\rho_s)$ is sampled independently for feature $p$, with $\rho_s=0.15$. The corrupted input is
$\widetilde x^{(M)}_{n,\ell}=\widetilde x_{n,\ell}\odot(1-M^s_{n,\ell})$.
The encoder and decoder are
\begin{equation}
\begin{aligned}
z^s_{n,\ell}&=f_{\theta_s}(\widetilde x^{(M)}_{n,\ell}),&
\widehat x_{n,\ell}&=g_{\theta_s}(z^s_{n,\ell}),\\
f_{\theta_s}:&\ 5\rightarrow64\rightarrow32\rightarrow4,&
g_{\theta_s}:&\ 4\rightarrow32\rightarrow64\rightarrow5 .
\end{aligned}
\label{eq:state_ae}
\end{equation}
The hidden layers use $\operatorname{GELU}(a)=a\Phi(a)$, where $\Phi$ is the standard-normal cumulative distribution function. Its smooth response retains weak signed activations from standardized continuous features. The per-sample objective is
\begin{equation}
\mathcal L_s=
\frac{\|M^s_{n,\ell}\odot(\widehat x_{n,\ell}-\widetilde x_{n,\ell})\|_2^2}
{\max(\|M^s_{n,\ell}\|_1,1)}
+\frac{0.1}{5}\|\widehat x_{n,\ell}-\widetilde x_{n,\ell}\|_2^2 .
\label{eq:state_mae_loss}
\end{equation}
The first term infers masked coordinates from the remaining variables; the second preserves complete-profile reconstruction. After training, unmasked states form deterministic representations for OCSVM fitting and inference. The signed residual and residual--latent representation are
\begin{equation}
\varepsilon^s_{n,\ell}=\widetilde x_{n,\ell}-\widehat x_{n,\ell},\quad
h^s_{n,\ell}=[(\varepsilon^s_{n,\ell})^\top,(z^s_{n,\ell})^\top]^\top
\in\mathbb R^9 .
\label{eq:state_representation}
\end{equation}

\subsubsection{RBF one-class state boundary}
A second training-only scaler produces $\bar h^s_{n,\ell}$. The RBF kernel and OCSVM decision function are
\begin{equation}
\begin{aligned}
K(a,b)&=\exp[-\gamma\|a-b\|_2^2],\\
d_{\mathrm{OCSVM}}(\bar h^s)
&=\sum_{q\in\mathrm{SV}}\alpha_qK(\bar h^s_q,\bar h^s)-b_{\mathrm{svm}},
\end{aligned}
\label{eq:state_ocsvm}
\end{equation}
where $\mathrm{SV}$ is the support-vector set, $\alpha_q$ are learned dual weights, and $b_{\mathrm{svm}}$ is the intercept. Here $\nu$ controls tolerated training violations and $\gamma$ controls RBF locality; \texttt{scale} is variance based. The selected head uses $\nu=0.05$ and $\gamma=\texttt{scale}$. Because positive OCSVM distance denotes normal support, the raw state anomaly score reverses the sign:
\begin{equation}
r_{\mathrm{state}}(n,\ell)=-d_{\mathrm{OCSVM}}(\bar h^s_{n,\ell}).
\label{eq:state_raw_score}
\end{equation}
The RBF boundary can follow a nonlinear, nonelliptical normal-state region without requiring an accurate density for every normal charging mode.

\subsubsection{Transition masked autoencoder and shrinkage distance}
Only the compact transitions in Eq. \eqref{eq:compact_transition} from observed normal user updates fit this branch. A transition scaler gives $\widetilde\delta_{n,\ell}=\mathcal S_t(\delta_{n,\ell})$. With masking probability $\rho_t=0.25$, the transition AE uses:
\begin{equation}
\begin{aligned}
z^t_{n,\ell}&=f_{\theta_t}(\widetilde\delta^{(M)}_{n,\ell}),&
\widehat\delta_{n,\ell}&=g_{\theta_t}(z^t_{n,\ell}),\\
f_{\theta_t}:&\ 5\rightarrow12\rightarrow6\rightarrow2,&
g_{\theta_t}:&\ 2\rightarrow6\rightarrow12\rightarrow5 ,
\end{aligned}
\label{eq:transition_ae}
\end{equation}
and the same masked/full reconstruction form as Eq.~\eqref{eq:state_mae_loss}. After training, unmasked transitions form deterministic representations for covariance fitting and inference:
\begin{equation}
\varepsilon^t_{n,\ell}=\widetilde\delta_{n,\ell}-\widehat\delta_{n,\ell}, ~~~
h^t_{n,\ell}=[(\varepsilon^t_{n,\ell})^\top,(z^t_{n,\ell})^\top]^\top
\in\mathbb R^7 .
\label{eq:transition_representation}
\end{equation}
Let $\bar h^t$ denote the standardized transition representations, $\widehat\mu_t$ their normal-update mean, and $S_t$ their sample covariance. Because the observed-update sample is small, direct inversion of $S_t$ can amplify estimation noise. Ledoit--Wolf shrinkage estimates
\begin{equation}
\widehat\Sigma_{\mathrm{LW}}
=(1-\lambda_{\mathrm{LW}})S_t+
\lambda_{\mathrm{LW}}\frac{\operatorname{tr}(S_t)}{7}I_7,
\label{eq:lw_covariance}
\end{equation}
where $I_7$ is the $7\times7$ identity and $\lambda_{\mathrm{LW}}\in[0,1]$ is estimated from the normal-update data. Shrinkage blends the noisy sample covariance with a stable spherical covariance before inversion. The raw transition score is the squared Mahalanobis distance.
\begin{equation}
r_{\mathrm{tr}}(n,\ell)=
(\bar h^t_{n,\ell}-\widehat\mu_t)^\top
\widehat\Sigma_{\mathrm{LW}}^{-1}
(\bar h^t_{n,\ell}-\widehat\mu_t).
\label{eq:transition_distance}
\end{equation}
It measures both change magnitude and whether demand, rate, timing, and duration move in a normally correlated direction.
Masking does not synthesize updates; its targets remain observed transitions. Both AEs use AdamW with learning rate $10^{-3}$, weight decay $10^{-5}$, batch size $\min(96,N_{\mathrm{tr}})$, and 180/160 state/transition epochs.

\subsubsection{Continuous score fusion and decision}
The two raw heads have different units and sample populations. The state reference contains all training-normal states, whereas the transition reference contains only observed training updates. Each score is standardized against its corresponding reference:
\begin{equation}
\begin{aligned}
s_{\mathrm{state}}&=\frac{r_{\mathrm{state}}-\mu_{r,s}}{\sigma_{r,s}+10^{-8}},&
s_{\mathrm{tr}}&=\frac{r_{\mathrm{tr}}-\mu_{r,t}}{\sigma_{r,t}+10^{-8}}.
\end{aligned}
\label{eq:head_standardization}
\end{equation}
Here $\mu_{r,s/t}$ and $\sigma_{r,s/t}$ are the corresponding training-reference means and standard deviations, respectively.
Equation~\eqref{eq:dual_score} then applies the selected low transition weight $\omega_{\mathrm{tr}}=0.15$. This is a signed correction rather than a second hard alarm: a normal-looking transition can reduce the score of a rare but legitimate revised state, while an unusual transition can promote a manipulated state that remains individually plausible. The state branch still detects abnormal activations and strong isolated states; the transition branch resolves ambiguity only when a request actually changes. The final decision is
\begin{equation}
\widehat y_{n,\ell}=\mathbf 1[
s_{\mathrm{dual}}(n,\ell)\ge s_{\mathrm{thr}}],
\label{eq:dual_decision}
\end{equation}
where disjoint normal calibration chooses the smallest threshold satisfying an overall true-negative rate (TNR) of at least 0.95 and an observed-update TNR of at least 0.90. Algorithm~\ref{alg:dual_branch} summarizes normal-only fitting and event-level inference.

\begin{algorithm}[thb!]
\caption{Dual-Branch Masked-AE Transition Boost}
\label{alg:dual_branch}
\begin{algorithmic}[1]
\Require Training normal states and observed normal transitions.
\State Fit $\mathcal S_s$; train the state masked AE by~\eqref{eq:state_mae_loss}.
\State Form $h^s$; fit its scaler and RBF OCSVM on normal states.
\State Construct~\eqref{eq:compact_transition} for observed updates and fit $\mathcal S_t$.
\State Train the transition masked AE; form $h^t$.
\State Fit the transition scaler, $\widehat\mu_t$, and $\widehat\Sigma_{\mathrm{LW}}$.
\State Record training-reference moments for both raw scores.
\For{each activation, observed update, or generated attack}
 \State Compute $s_{\mathrm{state}}$ and, when $I^{\mathrm{chg}}=1$, $s_{\mathrm{tr}}$.
 \State Fuse by~\eqref{eq:dual_score}; classify by~\eqref{eq:dual_decision}.
\EndFor
\end{algorithmic}
\end{algorithm}

\subsection{Source-Grouped CV and Calibration}
Seed 66 partitions the 1,213 sources into 970 development and 243 final-test sources while stratifying update type. Five source-grouped folds use 776 training and 194 validation sources each. Their validation update counts are 7, 7, 11, 8, and 9. The unified search evaluates 89 configurations in 445 fold tasks; every family is defined before final testing. For every fold and candidate, normal validation scores choose the most sensitive threshold satisfying the TNR constraints.
\begin{equation}
\mathrm{TNR}_{\mathrm{all,val}}\ge0.95,\qquad
\mathrm{TNR}_{\mathrm{update,val}}\ge0.90.
\label{eq:session_constraints}
\end{equation}
``Most sensitive'' means the lowest threshold that still attains both normal-acceptance rates. The second constraint makes benign-update discrimination an explicit selection requirement. Fixed validation attacks compute macro-F1 across twelve settings, with each setting contributing one twelfth regardless of its eligible attack count. To reward both average detection and stability across source partitions, the objective penalizes the foldwise standard deviation (SD), and configuration $\mathfrak c$ is ranked by
\begin{equation}
J_{\mathrm{cv}}(\mathfrak c)=\overline{\mathrm{F1}}_{\mathrm{macro}}(\mathfrak c)
-\lambda_{\mathrm{std}}\operatorname{SD}_{\mathrm{fold}}
[\mathrm{F1}_{\mathrm{macro}}(\mathfrak c)], \quad \lambda_{\mathrm{std}}=0.5,
\label{eq:robust_cv}
\end{equation}
with update TNR and overall TNR as tie breakers. Algorithm~\ref{alg:cv} summarizes this selection protocol. The best configuration in each family/protocol pair is retained for comparison, and the largest $J_{\mathrm{cv}}$ selects the formal model. It is refitted on 776 sources (811 states), calibrated on 194 disjoint normal sources (201 states, seven updates), and evaluated on 243 untouched sources (252 states, nine updates) and 1,505 attacks.

\begin{algorithm}[thb!]
\caption{Five-fold multi-input model selection}
\label{alg:cv}
\begin{algorithmic}[1]
\For{each model, protocol, and parameter configuration}
 \For{each source-grouped fold}
  \State Fit on fold-training normal states only.
  \State Select $s_{\mathrm{thr}}$ under~\eqref{eq:session_constraints}.
  \State Evaluate the twelve source-bound validation attacks.
 \EndFor
 \State Compute $J_{\mathrm{cv}}$ in~\eqref{eq:robust_cv}.
\EndFor
\State Retain each family/protocol winner and the global winner.
\State Refit, calibrate on normal data, and test each selected model once.
\end{algorithmic}
\end{algorithm}

\subsection{Evaluation and Statistical Reporting}
Table~\ref{tab:source_roles} gives the exact source-state accounting. Roles are fixed by source before attack generation, and every role includes all qualifying inputs from each assigned source. Consequently, no later input or attack derived from a test source can enter fitting, validation, or threshold calibration.

\begin{table}[thb!]
\centering
\caption{Source, normal-state, and observed-update counts.}
\label{tab:source_roles}
\begin{tabular}{lrrr}
\toprule
\textbf{Role} & \textbf{Sources} & \textbf{States} & \textbf{Updates}\\
\midrule
Session final training & 776 & 811 & 35\\
Session calibration & 194 & 201 & 7\\
Final test & 243 & 252 & 9\\
\bottomrule
\end{tabular}
\end{table}

Using true-positive (TP), false-positive (FP), true-negative (TN), and false-negative (FN) counts, the reported metrics are precision, recall, F1, and false-positive rate (FPR).
\begin{equation}
\begin{aligned}
\mathrm{Prec}&=\mathrm{TP}/(\mathrm{TP}+\mathrm{FP}),\\
\mathrm{Rec}&=\mathrm{TP}/(\mathrm{TP}+\mathrm{FN}),\\
\mathrm{F1}&=2\,\mathrm{Prec}\,\mathrm{Rec}/(\mathrm{Prec}+\mathrm{Rec}),\\
\mathrm{FPR}&=\mathrm{FP}/(\mathrm{FP}+\mathrm{TN}).
\end{aligned}
\label{eq:metrics}
\end{equation}
Accuracy is omitted because attack and normal counts are imbalanced and false
alarms have a distinct operational cost. Observed-update FPR is reported
separately from activation FPR. For $K_{\mathrm W}$ successes among $N_{\mathrm W}$ trials and proportion
$\widehat\pi=K_{\mathrm W}/N_{\mathrm W}$, the Wilson
confidence interval (CI) is used because update and attack-origin counts are small:
\begin{equation}
\begin{aligned}
&\mathrm{CI}_{\mathrm W}^{\pm}
={}\left[\widehat\pi+\frac{\zeta_{0.975}^2}{2N_{\mathrm W}}\right.\\
&\left.\quad\pm\zeta_{0.975}
\sqrt{\frac{\widehat\pi(1-\widehat\pi)}{N_{\mathrm W}}
+\frac{\zeta_{0.975}^2}{4N_{\mathrm W}^2}}\right]
\bigg/\left(1+\frac{\zeta_{0.975}^2}{N_{\mathrm W}}\right).
\end{aligned}
\label{eq:wilson}
\end{equation}
with $\zeta_{0.975}=1.96$. Wilson intervals remain valid for small groups and observed rates of zero or one, so they are emphasized for update subgroups.

\section{Results}
\subsection{Session-Level Model Selection}
All 445 fold tasks satisfy Eq.~\eqref{eq:session_constraints}, and all 22 model families enter CV. Table~\ref{tab:session_results} reports the 17 families with test F1 of at least 0.60; the complete search remains in the reproducibility artifacts. This presentation cutoff does not affect selection. Dual-Branch Masked-AE Transition Boost has the largest robust CV score and is selected before test evaluation.

\begin{table*}[thb!]
\centering
\caption{Reported session-level models on 252 normal states and 1,505 attacks. All 22 families enter CV; the table retains test F1 $\geq 0.60$ for readability. Bold marks the CV-selected model.}
\label{tab:session_results}
\footnotesize\renewcommand{\arraystretch}{1.02}\setlength{\tabcolsep}{2.4pt}
\begin{tabular}{p{5.3cm}lcccccccc}
\toprule
\textbf{Model family} & \textbf{Protocol} & \textbf{CV F1} & \textbf{CV SD} & $J_{\mathrm{cv}}$ & \textbf{Prec.} & \textbf{Recall} & \textbf{Test F1} & \textbf{FPR} & \textbf{Update TNR}\\
\midrule
\textbf{Dual-Branch Masked-AE Transition Boost} & \textbf{Transition} & \textbf{.684} & \textbf{.026} & \textbf{.671} & \textbf{.989} & \textbf{.704} & \textbf{.823} & \textbf{.048} & \textbf{1.000}\\
Profile-Anchor Transition Boost & Transition & .678 & .051 & .653 & .992 & .650 & .785 & .032 & 1.000\\
Raw OCSVM & Profile & .672 & .051 & .646 & .991 & .618 & .761 & .032 & 1.000\\
Masked AE+OCSVM & Profile & .650 & .058 & .621 & .990 & .596 & .744 & .036 & 1.000\\
AE+NF & Profile & .593 & .068 & .559 & .989 & .538 & .697 & .036 & 1.000\\
Context OCSVM & Stratified & .600 & .096 & .553 & 1.000 & .483 & .651 & .000 & 1.000\\
DevNet & Transition & .584 & .070 & .549 & 1.000 & .477 & .646 & .000 & 1.000\\
Raw GMM & Transition & .607 & .133 & .541 & 1.000 & .478 & .647 & .000 & 1.000\\
Context Mahalanobis & Stratified & .572 & .066 & .539 & 1.000 & .486 & .654 & .000 & 1.000\\
Raw KNN & Profile & .571 & .065 & .539 & .987 & .538 & .696 & .044 & 1.000\\
AE-GMM & Profile & .590 & .129 & .525 & .983 & .532 & .690 & .056 & 1.000\\
Deep SVDD & Profile & .570 & .091 & .525 & .994 & .513 & .677 & .020 & 1.000\\
AE-Mahalanobis & Transition & .568 & .097 & .519 & 1.000 & .477 & .646 & .000 & 1.000\\
Context KNN & Stratified & .536 & .035 & .518 & 1.000 & .481 & .650 & .000 & 1.000\\
Raw Mahalanobis & Transition & .553 & .070 & .517 & 1.000 & .477 & .646 & .000 & 1.000\\
Raw LOF & Transition & .513 & .014 & .506 & 1.000 & .477 & .646 & .000 & 1.000\\
Station-Conditioned Masked AE+OCSVM & Stratified & .516 & .088 & .472 & .976 & .452 & .618 & .067 & 1.000\\
\bottomrule
\end{tabular}
\end{table*}

\begin{figure*}[thb!]
\centering
\includegraphics[width=.7\textwidth]{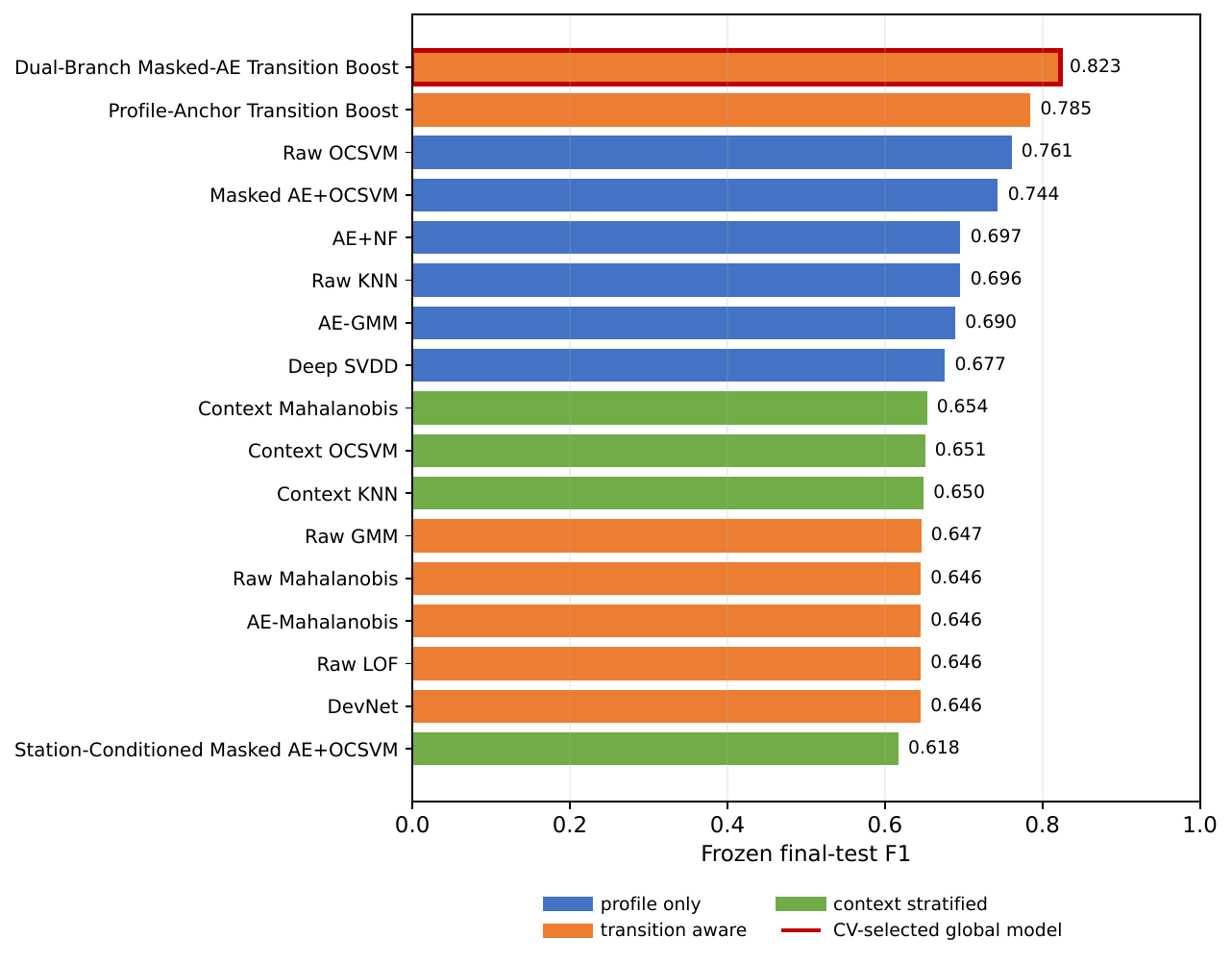}
\caption{Test F1 for the reported session-level model families with F1 $\geq 0.60$; colors denote feature protocol. Complete results are available in the reproducibility artifacts.}
\label{fig:session_f1}
\end{figure*}

Figure~\ref{fig:session_f1} visualizes the test F1 column of Table~\ref{tab:session_results}. The outlined Dual-Branch model leads with F1 0.823, followed by Profile-Anchor Transition Boost at 0.785 and raw OCSVM at 0.761. Blue, orange, and green denote profile-only, transition-aware, and context-stratified protocols. Model selection uses $J_{\mathrm{cv}}$, not this ranking.

The selected configuration in Fig.~\ref{fig:dual_branch_architecture} uses state widths $(64,32)$, latent dimension 4, $\rho_s=0.15$, and an RBF OCSVM with $\nu=0.05$ and $\gamma=\texttt{scale}$; its transition AE uses widths $(12,6)$, latent dimension 2, $\rho_t=0.25$, Ledoit--Wolf distance, and $\omega_{\mathrm{tr}}=0.15$. Normal-only calibration gives $s_{\mathrm{thr}}=1.43054$, overall TNR 0.9652, and update TNR 1.000. Final testing yields $(\mathrm{TN},\mathrm{FP},\mathrm{FN},\mathrm{TP})=(240,12,445,1060)$. Its $J_{\mathrm{cv}}=0.671$ exceeds Profile-Anchor Transition Boost's 0.653 and determines the formal selection.

\subsection{Observed-Update Discrimination}
Table~\ref{tab:update_results} isolates the central benign-update requirement. Across the five fold-specific validation thresholds, the selected dual-branch model accepts 41/42 held-out updates (TNR 0.976; Wilson 95\% CI $[0.877,0.996]$). Its refitted final model accepts every test update, including all demand-only, departure-only, and joint changes; the aggregate 9/9 interval
is $[0.701,1.000]$. These intervals prevent the small event count from being mistaken for a precise population estimate.

\begin{table}[thb!]
\centering
\caption{Formal dual-branch model behavior on observed updates and attack origins. The rate is TNR for benign rows and recall for attack rows.}
\label{tab:update_results}
\footnotesize\renewcommand{\arraystretch}{1.04}\setlength{\tabcolsep}{3.2pt}
\begin{tabular}{lrrr}
\toprule
\textbf{Group} & \textbf{Count} & \textbf{Correct} & \textbf{Rate}\\
\midrule
CV benign updates (out of fold) & 42 & 41 & .976\\
Test demand updates & 2 & 2 & 1.000\\
Test departure updates & 3 & 3 & 1.000\\
Test joint updates & 4 & 4 & 1.000\\
Activation-origin attacks & 1,479 & 1,042 & .705\\
Update-origin attacks & 26 & 18 & .692\\
\bottomrule
\end{tabular}
\end{table}

Attacks launched from observed-update and activation states have recalls 0.692 (Wilson 95\% CI $[0.500,0.835]$) and 0.705 ($[0.681,0.727]$), respectively. The close rates show comparable coverage for attacks originating before and after
a benign update without treating the updates themselves as anomalies. The small 26-attack subgroup nevertheless precludes a strong equivalence claim.

\subsection{Attack-Specific Behavior}
\begin{figure}[thb!]
\centering
\includegraphics[width=\linewidth]{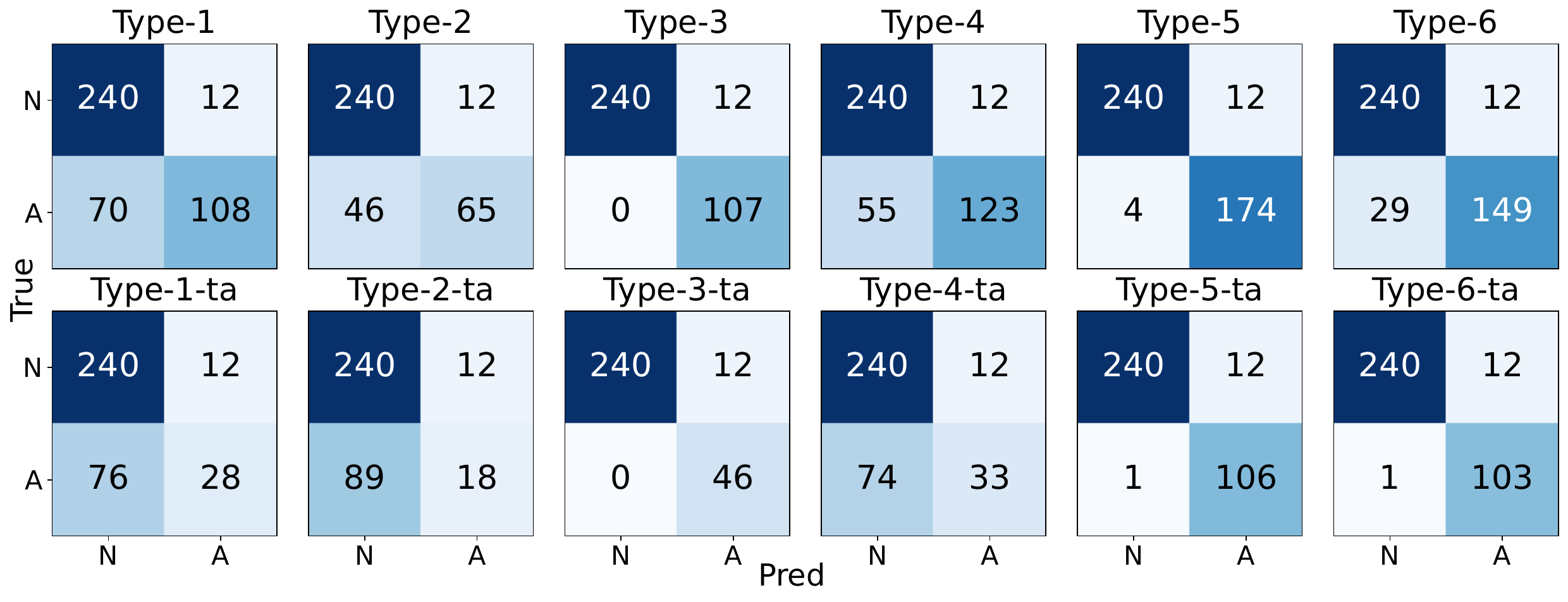}
\caption{Formal Dual-Branch Masked-AE Transition Boost confusion matrices for the twelve final CMA settings.}
\label{fig:formal_cm}
\end{figure}

Figure~\ref{fig:formal_cm} reports the formal dual-branch model; complete metrics are released with the per-attack results. Each panel uses the same 252 normal states. Type-1-ta, Type-2-ta, and Type-4-ta are difficult partly because their per-profile perturbations shrink near requested departure: $\Delta d_i^{\mathrm{att}}\propto H_n(ta)$ in~\eqref{eq:attack_state}, and Table~\ref{tab:cma_types} likewise scales their timing changes with $H_n(ta)$. Thus, a coordinated attack can alter aggregate load while leaving late-session profiles near normal support. Types 5-ta and 6-ta perturb demand and timing more clearly and reach recalls 0.991 and 0.990. Of 1,505 attacks, 26 originate from observed-update states and 18 are detected.

\section{Discussion}
Three findings follow. First, recorded user updates must be modeled as legitimate transitions. Source grouping, update-TNR constraints, and update-origin reporting prevent recall gains obtained by rejecting authorized revisions. The selected model accepts 41/42 held-out development updates and all nine test updates, with similar recall for activation- and update-origin attacks.

Second, the branches are complementary: the state masked AE and OCSVM learn support for initial and revised requests, while the transition masked AE and shrinkage covariance head learn legitimate demand--timing changes. Gated fusion adjusts ambiguous revised states without removing the state-only activation path. Its CV gain over profile anchor and raw OCSVM is encouraging, although 42 development updates limit conclusions about larger transition heads.

Third, attack physics determines separability. Per-attack analysis explains the resulting failure modes but never selects a detector after observing attack type.

\section{Limitations and Deployment Considerations}
The CMAs are physically motivated simulations rather than observed intrusions. The fixed $C_{\mathrm{att}}=0.15$ represents one impact--stealth point; adaptive adversaries may use weaker, targeted, or detector-aware changes, so the results support only the twelve defined settings.

The data expose 51 qualifying updates from 45 sessions, with nine updates and 26 update-origin attacks in the final test. ACN may omit app-level edits not retained in its input history, and the resulting wide Wilson intervals mean that observed update performance is not a precise population-wide rate. Broader sites, dates, and application logs are needed to assess transition-aware generalization.

The compact features omit charger rating, vehicle capability, user history, tariffs, feeder state, and power telemetry. Fitting is normal-only, but simulated validation attacks select configurations; the framework is therefore simulation-assisted rather than fully label-free. Thresholds require periodic recalibration as behavior changes. Detection remains decision support: a CSMS must balance grid risk and user inconvenience when confirming, freezing, or escalating an edit.

\section{Conclusion}
This paper benchmarked cyberattack detection for multi-input EV charging while treating recorded user revisions as normal. The source-bound design applies six physical CMAs and six coordinated variants under leakage-controlled roles. Source-grouped cross-validation selects Dual-Branch Masked-AE Transition Boost, which separates current-state support from normal transition geometry. It achieves F1 0.8227 on 1,505 final attacks, accepts all nine test updates, and gives comparable recall for activation- and update-origin attacks. Reliable CMA detection therefore requires authorized revisions to be modeled throughout splitting, learning, selection, and calibration.

Potential future work will expand observed updates across sites and seasons, add charger and vehicle context, and test adaptive attacks. A completed causal sliding-window fusion extension addresses synchronized multi-session attacks by combining cross-session temporal evidence with local anomaly scores; details are reserved for a separate study.

\section*{Acknowledgments}
This work was supported by National Science Foundation under grant DMS-2229417.

\section*{Code and Data Availability} 
The code and dataset repository is available at: \url{https://github.com/Hannan-Chen/Cyberattack-Detection-in-EV-Charging-Systems}.

\section*{Declaration of AI-Assisted Research} 
Generative AI assisted coding, visualization, and writing; the authors verified all AI-assisted content and take full responsibility for the work.

\bibliographystyle{./bibliography/IEEEtran}
\bibliography{./refs}
\end{document}